\input epsf
\documentstyle[12pt]{article}
\begin{document}
\rightline{Preprint YERPHI-1470(7)-96}
\vspace{1cm}
\begin{center}
{\Large HIGGS BOSON MASSES IN NMSSM WITH SPONTANEOUS CP-VIOLATION}  \\
\vspace{1cm}
{\large ASATRIAN H.M., YEGHIYAN G. K.} \\

\vspace{1cm}
{\em Yerevan Physics Institute, Alikhanyan Br. 2, Yerevan, Armenia}\\
{\em e-mail: "asatryan@vx1.yerphi.am"}\\
\end{center}
\vspace{5mm}

\centerline{{\bf{Abstract}}}
 The Higgs boson mass problem is considered in the next-to-minimal
supersymmetric standard model for the case of the spontaneous CP
violation. The renormalization group equations for
the gauge, Yukawa and scalar coupling constants,
the effective Higgs potential and lower experimental
bounds on   Higgs boson and chargino masses are analyzed.
The restrictions on the Higgs boson masses are found.
\vspace{2cm}

1.The aim of our paper is to consider the problem of Higgs
boson masses in the next to minimal supersymmetric standard model (NMSSM)
for the case of the presence of spontaneous CP violation.
Such a model contains an additional Higgs singlet, as compared
with the minimal supersymmetric standard model (MSSM) \cite{1}.
It is known, that while in MSSM one can't realize the realistic
scenario with spontaneous CP-violation \cite{2}, in NMSSM
such a scenario, generally speaking, is realizable \cite{3,4}.
However, experimental bounds on the Higgs masses strongly
restrict the space of parameters of theory, where such a scenario
could be realized. In this paper we will continue the consideration
of the Higgs boson masses problem for the case with
spontaneous CP violation in NMSSM, taking into account the existing
experimental restrictions on the Higgs boson and chargino masses and
using the renormalization group equations and Higgs effective potential
analysis. This allows us to obtain more restrictive bounds on Higgs boson
masses than those in Ref. \cite{3}.

2.The Higgs sector of NMSSM consists of two Higgs doublets
$H_{1} =
\left(
\begin{array}{c}
\xi_{1}^{+} \\
\xi_{1}^{o}
\end{array} \right)$,
$H_{2} =
\left(
\begin{array}{c}
\xi_{2}^{o} \\
\xi_{2}^{-}
\end{array} \right)$
with hypercharges $Y(H_{1})=1, Y(H_{2})=-1$ and the complex
$SU(2)_{L} \times U(1)_Y$ singlet N. The superpotential for
one quark generation is the following
\cite{5}:
\begin{equation}
W = \frac{\lambda_{1}}{3}N^{3} + \lambda_{2}NH_{1}H_{2} +
    h_{u}H_{2}Q_{L}^{c}u_{R} +
h_{d}H_{1}Q_{L}^{c}d_{R}                                       
\end{equation}
where
\begin{displaymath}
Q_{L} =
\left(
\begin{array}{c}
u_{L} \\
d_{L}
\end{array} \right) \\
\hspace{0.5cm}
Q_{L}^{c} = i\sigma_{2}Q_{L}^{*}
\end{displaymath}
The scalar potential for the Higgs fields is the
following \cite{3,5,7,8}
\begin{eqnarray}
\nonumber
V & = & \frac{1}{2}a_{1}(H_{1}^{+}H_{1})^{2} +
\frac{1}{2}a_{2}(H_{2}^{+}H_{2})^{2}  +
a_{3}|H_{1}|^{2}|H_{2}|^{2}+ a_{4}|H_{1}H_{2}|^{2} +
a_{5}|N|^{2}|H_{1}|^{2}   \\
& + & a_{6}|N|^{2}|H_{2}|^{2} +
a_{7}((N^{2})^{+}H_{1}H_{2} + h.c.) + a_{8}|N|^{4} +    
+m_{4}(NH_{1}H_{2}   \\
\nonumber
& + & h.c.)   + \frac{m_{5}}{3}(N^{3} + h.c) +
m_{1}^{2}|H_{1}|^{2} +
m_{2}^{2}|H_{2}|^{2} + m_{3}^{2}|N|^{2}
\end{eqnarray}
The parameters of the potential (2) are connected by the relations
(for the more detailed discussion of the results, represented in this
chapter, see \cite{3}):
\begin{eqnarray}
\nonumber
a_{5} & = & a_{6} = \lambda_{2}^{2}, \hspace{0.3cm}
a_{7} = \lambda_{1}\lambda_{2},  \hspace{0.3cm}
a_{8} = \lambda_{1}^{2}   \\
a_{1} & = & a_{2} = \frac{1}{4}(g_{1}^{2}+g_{2}^{2}),  \hspace{0.2cm}
a_{3} = \frac{1}{4}(g_{2}^{2}-g_{1}^{2}),  \hspace{0.2cm}
a_{4} = \lambda_{2}^{2}-\frac{1}{2}g_{2}^{2}     
\end{eqnarray}
where $g_{1}$, $g_{2}$ are gauge coupling constants of the gauge
groups $U(1)_{Y}$ and $SU(2)_{L}$ respectively. These relations are
valid for the supersymmetry breaking scale $M_{s}$ and higher.
Below the supersymmetry
breaking scale the energy behavior of the coupling constants
$a_{i} (i=1,2,...,8)$
(and also of the Yukawa and gauge couplings) is given by
renormalization group equations.
The mass
parameters of the potential (2) $m_{j}$, j=1,..,5 are connected
with the supersymmetry breaking.
Unlike Refs \cite{5,6,7,8} we consider the model, where supersymmetry
breaking terms are not universal \cite{4,9}. This means that the
parameters $m_{j}$ are independent.

If the Higgs fields $H_{1}$, $H_{2}$, $N$ in potential (2) develop
nonzero VEV's, the electroweak symmetry breaking takes place.
To provide for the electric charge conservation we
choose these VEV's in the following form:
\begin{displaymath}
<H_{1}> =
\left(
\begin{array}{c}
0 \\
v_{1}
\end{array} \right) \\
\hspace{0.5cm}
<H_{2}> =
\left(
\begin{array}{c}
v_{2}e^{i\varphi} \\
0
\end{array} \right)
\hspace{0.5cm}
<N> = v_{3}e^{i\alpha}
\end{displaymath}
The case of the spontaneous CP-violation corresponds to vacuum
with nonzero phases: $\varphi \neq 0$ $\alpha \neq 0$ .

After the spontaneous
breaking of electroweak symmetry  five neutral and one complex
charged scalar fields appear. Excluding  the Goldstone mode, we
obtain $5 \times 5$ symmetric mass matrix for neutral fields
$\Phi_{1}$, $\Phi_{2}$, A,
$N_{1}$, $N_{2}$, where CP-even fields $\Phi_{1}$, $\Phi_{2}$ and
CP-odd field A are contained by Higgs doublets $H_{1}$, $H_{2}$ and
CP-even field $N_{1}$ and CP-odd field $N_{2}$ are contained by the
singlet N.
The mass of charged Higgs is given by:
\begin{equation}
m^{2}_{H^{+}} = - \frac{2(m_{4}cos(\varphi + \alpha) + a_{7}v_{3}cos(
\varphi - 2\alpha))v_{3}}{sin2\beta} - a_{4}\eta^{2}                  
\end{equation}
where
\begin{equation}
\eta^{2} = v_{1}^{2} + v_{2}^{2} = (174GeV)^{2},                      
\end{equation}
and
\begin{equation}
1 < \tan\beta = \frac{v_{2}}{v_{1}} = \frac{h_{b}}{h_{t}}           
\frac{m_{t}}{m_{b}} < 60
\end{equation}
where \cite{10,11} $m_{t}=(175 \pm 15)GeV$,
$m_{b}=(3.5 \pm 0.5) GeV$ are t- and b-quark masses respectively
and  $1 \leq h_{t} \leq 1.1$, $h_{b} \leq 1.1$ are their
Yukava coupling constants.

To obtain the restrictions on Higgs particles masses,
we have  investigated the restrictions on parameters
of the Higgs potential (2).
The restrictions for coupling constants $a_{i}, i=1,...8$
can be obtained, analyzing the renormalization group equations
for $a_{i}, i=1,...8$, gauge $g_{r}, r=1,2,3$ and t- and b-quark
Yukawa couplings
$h_{t}$, $h_{b}$
(one can neglect other Yukawa couplings because of their smallness with
regard to $h_{t}$ and $h_{b}$) in the region between electroweak breaking
scale and the supersymmetry breaking scale and from the supersymmetry
breaking scale up to unification scales ($\sim (10^{16}-10^{18})GeV$).
We assume that all of the ( gauge, Yukawa, scalar) coupling constants
of our theory are small between weak and
unification scales, so that the
perturbation theory applies \cite{3,12,13}. Besides this, for the theory to be
a correct one, the condition of vacuum stability is necessary.
These conditions give some restrictions for values
of scalar coupling constants
$a_{i}$ at $Q=M_{Z}$ .
The Higgs VEV's $v_{1}, v_{2}$ are restricted by conditions
(5),(6).
The mass parameters $m_{j}$, j=1,...,5 are connected with the
supersymmetry breaking and are of the order of supersymmetry breaking
scale or smaller. The restrictions on parameters of theory are also
obtained from minimum conditions on the Higgs potential (2)  and
from the requirement of positiveness of Higgs boson masses squared.
Using above mentioned restrictions on parameters of theory, we obtain
that in the case of the absence of spontaneous CP-violation the lightest
neutral Higgs boson has the mass of order of $\eta$ or smaller.
The mass of this particle depends on the radiative corrections to the
scalar coupling constants: due to these corrections one increases about
40-60 GeV , depending on supersymmetry breaking scale.
The
remaining Higgs particles, however, can be as heavy as the supersymmetry
breaking scale. This situation is similar to one in the minimal
supersymmetric standard model. In spite of the MSSM has smaller number of
independent parameters, only for the lightest Higgs boson an upper
bound on the mass can be obtained. It is obtained that this particle
always lighter
than Z-boson, if the tree-level potential is analyzed. However,
due to the radiative corrections to the MSSM Higgs potential the mass of
the lightest Higgs boson increases significantly (see for example
\cite{6} and references therein).

In the case of the presence of spontaneous CP-violation
situation is much more interesting.
The requirement of positiveness of Higgs particles masses
squared can be satisfied, only if the supersymmetry breaking scale is
higher than 100GeV.
In other words, we
come to conclusion that the scenario with spontaneous CP-violation
can be realized, only if the supersymmetry breaking scale is higher
than 100GeV.  In this case at least three neutral detectable
(nonsinglet) Higgs
particles exist
with masses of the order of $\eta$ or smaller. The charged Higgs boson also
has the mass of this order. As for the two remaining neutral Higgs
bosons, masses of these ones can be as large as the supersymmetry breaking
scale. However, two heaviest neutral Higgses must be almost
$SU(2)_{L} \times U(1)_{Y}$ singlets, if they have masses much larger
than $\eta$.

Such a difference between the cases of the presence and the absence of
spontaneous CP-violation takes place due to the additional restriction
on parameters of theory in the case of the presence of spontaneous
CP-violation, which follows from the minimum conditions for nonzero
vacuum phases . This restriction, combined with the requirement
of positiveness of Higgs particles masses squared, leads to the
following restriction on parameters of theory
\begin{eqnarray}
\nonumber
max((a_{3} + a_{4} - \sqrt{a_{1}a_{2}})\frac{\eta^{2}\sin2\beta}{2v_{3}^{2}}
,\frac{a_{7}^{2}\eta^{2}\sin2\beta}{2a_{8}v_{3}^{2}})<       \\
< -a_{7}\frac{\sin3\alpha}{\sin(\varphi+\alpha)} < (a_{3} + a_{4} +  
\sqrt{a_{1}a_{2}}) \frac{\eta^{2} \sin2\beta}{2v_{3}^{2}}
\end{eqnarray}
which in his turn leads to results, described above

3. Besides this, we have found in \cite{3}
that in the case of the presence of spontaneous CP-violation
the lightest neutral Higgs
(detectable or no) has the mass
\begin{equation}
m_{h} \leq 35GeV                    
\end{equation}
On the
other hand the experimental restrictions on Higgs boson masses
in the minimal supersymmetric standard model are the following
\cite{14,15}:
\begin{equation}
m_{h} > 45GeV, \hspace{2cm}        
m_{H^{+}} > 45GeV
\end{equation}
for CP-even and charged Higgses respectively and
\begin{eqnarray}
\nonumber
m_{A} & > & 27 GeV \hspace{0.5cm} for \hspace{0.5cm}
1 < \tan\beta < 1.5, \\
m_{A} & > & 45 GeV \hspace{0.5cm} for \hspace{0.5cm}          
\tan\beta > 1.5
\end{eqnarray}
for the CP-odd Higgs A.
Unlike the MSSM, NMSSM contains an additional complex
$SU_{L}(2) \times U_{Y}(1)$ singlet field N so that in NMSSM with the
presence of spontaneous CP-violation the lightest neutral Higgs,
generally speaking, is some
mixing of five real scalar fields: two CP-even and one CP-odd fields,
contained by Higgs doublets $H_{1}$ and $H_{2}$, and two singlet fields
$N_{1}$,$N_{2}$. This fact and conditions (8)-(10) lead us to the
following conclusion.

The scenario with spontaneous CP-violation in NMSSM, generally speaking,
can be realistic for $\tan\beta > 1$, if the lightest neutral Higgs boson
is almost $SU_{L}(2) \times U_{Y}(1)$ singlet. Also
the scenario with spontaneous CP-violation
for  $1 < \tan\beta <1.5$ can be realistic, if the lightest neutral Higgs
is some mixing of the CP-odd field A and singlet fields $N_{1}$ and
$N_{2}$. However, the last scenario seems to be improbable. First of
all, the new experiments soon can exclude also the possibility of
existing of neutral Higgs with the mass smaller than 35GeV. Besides
this, the region of $1 < \tan\beta < 1.5$ is permissible only for
$m_{t} \approx 160GeV$, i.e. for the minimal value of t-quark mass, allowed
by \cite{10}. The new experimental data \cite{16} give the average value
of t-quark mass $m_{t}=(180 \pm 12)GeV$. For a latter values of
t-quark mass the allowable region of $\tan\beta$ is $1.5 < \tan\beta < 60$.
This means that the lightest neutral Higgs must be almost
$SU(2)_{L} \times U(1)_{Y}$ singlet, to avoid the contradiction with
experiment.
As it was mentioned above, in this case
three detectable (i.e. nonsinglet) neutral Higgs
particles with masses of the order of $\eta$
or smaller always exist.

4.Thus we must investigate the case, when
(in the case of the presence of spontaneous CP-violation) the
lightest neutral Higgs is almost singlet. The spectrum of the neutral
Higgs masses is described by
the $5 \times 5$ mass matrix $M^{2}$, given
in \cite{3} by formula (A1) in Appendix A. This matrix was obtained
for the vector H, which has the form $H^{T} = (\Phi_{1}, \Phi_{2}, A,
N_{1}, N_{2})$. The lightest neutral Higgs boson is the some linear
combination of fields $\Phi_{1}$, $\Phi_{2}$, A,
$N_{1}$, $N_{2}$, i. e.
\begin{displaymath}
h=x_{1} \Phi_{1} + x_{2} \Phi_{2} + x_{3} A +
x_{4} N_{1} + x_{5} N_{2}
\end{displaymath}
so that the coefficients $x_{k}$, k=1,..,5 can be found from the
equation
\begin{equation}
M^{2} X_{1} = m^{2}_{h} X_{1}                                 
\end{equation}
where $X_{1}^{T}=(x_{1},x_{2},x_{3},x_{4},x_{5})$.
The requirement that the lightest neutral Higgs boson must be almost
$SU(2) \times U(1)$ singlet means that
the following condition must be satisfied:
\begin{equation}
\frac{x_{4}^{2} + x_{5}^{2}}{|X_{1}|^{2}} \rightarrow 1                
\end{equation}
The analytical investigation of equation (11) shows that at
least two cases exist,
when the condition (12) can be satisfied. These  cases are
the following:
\begin{eqnarray}
& a) & \hspace{0.5cm} a_{7}v_{3} \ll \eta \hspace{0.5cm} and
\hspace{0.5cm} v_{3} \gg \eta \sin2 \beta  \\                      
& b) &\hspace{0.5cm} \varphi \ll \alpha \ll 1 \hspace{0.5cm} and
\hspace{0.5cm} a_{7}v_{3}^{2} \sim \eta^{2} \sin2 \beta            
\end{eqnarray}
As one can observe, lower experimental bounds on the Higgs boson masses
give additional
restrictions on the singlet VEV $v_{3}$.
On the other hand, it is known that the singlet Higgs boson
interacts also with charged and neutral Higgs fermions (Higgsinos).
This means that the restrictions (13) and (14)
on $v_{3}$ can bring some new restrictions on chargino and neutralino
masses. In particular, it
can happen that in parameter space, satisfying the conditions (13)
and (14), charginos and neutralinos will have masses, which are
smaller than existing lower experimental bounds on their
values \cite{17}.
An analytical and numerical investigations show that while the lower
experimental bounds on neutralino masses don't give some essential
restrictions on the Higgs boson masses, the experimental bound on the
lightest chargino mass $m_{c_{1}} > 45 GeV$ plays crucial role. Therefore
in the next chapter we will consider the chargino masses problem in
more detail.

5.The method of finding chargino masses in
the minimal supersymmetric standard model is described in Ref.
\cite{18}. The difference between our case and the MSSM is that
we must make replacements $v_{2} \rightarrow v_{2} e^{i \varphi}$
and $\mu \rightarrow \lambda_{2} v_{3} e^{i \alpha}$, where
$\lambda_{2} \approx \lambda_{2}(M_{s})$.
Then the lightest chargino mass square is given by
\begin{eqnarray}
\nonumber
m^{2}_{C_{1}} & = &  \frac{1}{2} (M_{2}^{2} + \lambda_{2}^{2} v_{3}^{2} +
g_{2}^{2} \eta^{2} -
((M_{2}^{2} + \lambda_{2}^{2} v_{3}^{2} +                     
g_{2}^{2} \eta^{2})^{2} -
4 M_{2}^{2} \lambda_{2}^{2} v_{3}^{2} -  \\
& - & g_{2}^{4} \eta^{4} \sin^{2} 2 \beta
+ 4 M_{2} \lambda_{2} v_{3} g_{2}^{2} \eta^{2} \sin 2 \beta
\cos(\varphi + \alpha))^{1/2})
\end{eqnarray}
where $M_{2}$ is SU(2) gaugino
mass arising from supersymmetry breaking.

Figure 1 presents the minimal value of $\lambda_{2} v_{3}$ as a function
of $\tan \beta$ for which the condition
$m_{C_{1}} > 45GeV$ \cite{17} is  satisfied.
These restrictions on $\lambda_{2} v_{3}$
was found from  the numerical analysis of (15). As it follows
from Fig. 1,
\begin{equation}
\lambda_{2} v_{3} > 45GeV                  
\end{equation}
for $\tan \beta > 10$.
Thus we have shown that the lower experimental bound on the lightest
chargino mass gives new restriction on the singlet VEV $v_{3}$.

6.Let us now consider the cases a) and b) in more detail.
Before doing it we want to stress the following. It can be obtained
from the renormalization group equations analysis that
$a_{7} \approx \lambda_{1}(M_{s}) \lambda_{2} (M_{s})$.
Also analytical and numerical investigations of the mass matrix
$M^{2}$ show that the requirement of positiveness of Higgs boson masses
squared can't be satisfied, if $\lambda_{1} \ll \lambda_{2}$
(numerically $\lambda_{1} < 0.1 \lambda_{2}$). Taking into account
two above mentioned conditions, we proceed now to consideration of
cases a) and b). We obtain that
in the case
a) the conditions
\begin{equation}
a_{7} \ll 1, \hspace{0.5cm} \lambda_{2} \ll 1       
\end{equation}
must be
satisfied. Really, as it follows from (13), $v_{3} \gg \eta$
for small values of $\tan \beta$
so that the necessity of the condition (17) is obvious.
As for the large values of $\tan \beta$, then (17) is necessary to the
restriction (16) to be satisfied.
Consequently, we obtain in this case that two neutral
Higgses are almost singlets: one, the CP-odd Higgs boson $N_{2}$
has the mass of the order of few GeV or smaller, and the mass of the
CP-even Higgs $N_{1}$ can be as heavy as the supersymmetry breaking scale.
One of the detectable (nonsinglet) neutral Higgses
is almost CP-odd field A. Using the condition (7), we obtain that
 for the mass of A the following condition must be satisfied:
\begin{equation}
m_{A}^{2} < (a_{4} + a_{3} + \sqrt{a_{1}a_{2}}) \eta^{2}    
\end{equation}
If $\tan \beta < 20$, then from the results, obtained from the analysis
of renormalization groups equations for scalar coupling constants
$a_{1}$, $a_{2}$, $a_{3}$, $a_{4}$ for the case $\lambda_{2} \ll 1$
\cite{3}
we obtain that $m_{A} < 46GeV$
and $m_{A} < 52GeV$  for  $M_{s} \sim 1TeV$ and
$M_{s} \sim 10 TeV$  respectively.
In other words, the scenario with spontaneous
CP-violation, described above, is disfavored for $\tan \beta < 20$.
For larger values of  $\tan \beta$ the mass of A increases
due to increasing of $a_{1}$.
It is also necessary to stress
that for large values of $\tan \beta$ masses of two other
nonsinglet neutral Higgses are given by
\begin{equation}
m_{h_{1}} \approx m_{A} \hspace{0.5cm}
m_{h_{2}} \approx 2 a_{2} \eta^{2}                         
\end{equation}
and these particles are almost CP-even fields $\Phi_{2}$ and
$\Phi_{1}$ respectively. Last result (19)
can be obtained, putting
in nonsinglet part of the mass matrix $M^{2}$
$\sin 2 \beta \approx 0$ and $\cos 2 \beta \approx 1$.

As for the chargino and neutralino masses, they can be as heavy
as the supersymmetry breaking scale, if
$\lambda_{1}(M_{s})$,
$\lambda_{2}(M_{s})$, and
consequently $a_{7}$, are sufficiently small.

Notice also that in the case a) the condition $\varphi \gg \alpha$
must be satisfied, if $\tan \beta \gg 1$ or if $\lambda_{2} v_{3} \gg
\eta$.

In the case b) ($\varphi \ll \alpha \ll 1$) the conditions (14), (16) and
the requirement of positiveness of Higgs boson masses squared can't be
satisfied simultaneously. The case b) is valid only for small values
of $\tan \beta$, where
the condition (16) is not valid. In this case the mass of CP-odd Higgs
particle A is further restricted by the condition (18). However, as
the numerical analysis of this case shows, this particle,
generally speaking, can be heavier than 50GeV.
As in the case a) the lightest singlet Higgs is almost CP-odd
field $N_{2}$. Three remaining neutral Higgses are almost
some mixings of CP-even
fields $\Phi_{1}$, $\Phi_{2}$ and $N_{1}$.

7.As it could be seen from the previous analysis,
in this model the nonsinglet
Higgs boson masses strongly depend on the radiative corrections to
the scalar
coupling constants. The several approaches exist to take into account
radiative corrections to the Higgs boson masses. One of them is the
renormalization group equations analysis approach, which was used in
ref. \cite{3} and here up to this chapter. The other one is an effective
potential approach, where except of corrections to the coupling constants
the new terms, arising from the radiative corrections to the Higgs potential,
are also taken into account \cite{19,20,21}.
In spite of the last approach more perfectly
takes into account the radiative corrections, in NMSSM without spontaneous
CP-violation both of approaches give almost the same results (compare, for
example, the results of \cite{3,5} and \cite{6}). It is interesting to
investigate what does take place in the case of the presence of spontaneous
CP-violation. In this chapter we will consider an effective potential
approach, taking into account only t- and b- quark and squark one-loop
corrections because of the largeness of $h_{t}$ and (for large $\tan \beta$)
$h_{b}$ compared
with other couplings constants. This means that we take
into account only the
loops, containing the following vertexes:
\begin{eqnarray}
\nonumber
V_{tb}=h_{t}^{2}(|\tilde{Q}_{L}^{+} H_{2}|^{2} +
|H_{2}|^{2} |\tilde{t}_{R}|^{2})
+ h_{b}^{2}((|\tilde{Q}_{L}^{+} H_{1}|^{2} +
|H_{1}|^{2} |\tilde{b}_{R}|^{2}) +  \\
\nonumber
+h_{t} h_{b} (\tilde{t}_{R}^{+} H_{2}^{+} H_{1} \tilde{b}_{R} + h.c) +
 (h_{t} \bar{Q}_{L} H_{2} t_{R} +
h_{b} \bar{Q}_{L} H_{1} b_{R} + h.c)
\end{eqnarray}
However, we continue take into account
the radiative corrections, connected also with other coupling
constants, using the renormalization groups analysis approach, where
we rewrite the renormalization groups equations for the scalar
coupling constants $a_{i}$, i=1,...,8 (see the preprint version of
\cite{3}) without terms, containing only $h_{t}$ and $h_{b}$.
To calculate the scalar loops contributions we use the method,
described in Ref.
\cite{19}. Proceeding from supersymmetry, the contribution of quark loops
is the same as the contribution of respective squark loops, taken with
$m^{2}=0$, where $m^{2}$ is supersymmetry breaking squark masses
(for a simplicity we take them equal).
The method of finding the contribution of these loops
to the masses of Higgs bosons is also well-known (see, for example,
\cite{20} and references therein).
Using these methods, we must do the following. The restrictions on
coupling constants must be found
in the way described in the chapter 2. Then the transformations
\begin{eqnarray}
\nonumber
a_{1} & \rightarrow & a_{1} + \frac{3}{8 \pi^{2}} h_{b}^{4}
\hspace{0.1cm}
\ln (1 + \frac{m^{2}}{m_{b}^{2}})  \\
\nonumber
a_{2} & \rightarrow & a_{2} + \frac{3}{8 \pi^{2}} h_{t}^{4}
\hspace{0.1cm}
\ln (1 + \frac{m^{2}}{m_{t}^{2}})   \\                             
a_{3} & \rightarrow & a_{3} + \frac{3}{8 \pi^{2}} h_{b}^{2}
h_{t}^{2} \hspace{0.1cm}
\ln (1 + \frac{m^{2}}{m_{t}^{2}})  \\
\nonumber
a_{4} & \rightarrow & a_{4} - \frac{3}{8 \pi^{2}} h_{b}^{2}
h_{t}^{2} \hspace{0.1cm}
\ln (1 + \frac{m^{2}}{m_{t}^{2}})
\end{eqnarray}
in Higgs boson mass-matrices must be done. It is clear that after
this procedure the qualitative analysis of the restrictions
on the Higgs boson masses, done in ref. \cite{3},
generally speaking, remain true. The difference is only that
due to the  transformations (20) the requirement of positiveness
of the Higgs boson masses squared can be satisfied
now also for $M_{s} \sim 100
GeV$, so that the scenario with spontaneous CP-violation can be
realized also at this scale.  For $M_{s} \sim 1 TeV$ and
$M_{s} \sim 10 TeV$ the following quantitative changes of the
restrictions on the Higgs boson masses take place. In spite of
$m_{t}^{2} \sim M_{Z}^{2}$, due to the fact that
$h_{t}(Q^{2}) < h_{t}(M_{Z})^{2})$, if $Q^{2} > M_{Z}^{2}$
(this inequality is obtained from the renormalization group
analysis for $h_{t}$) and, consequently,
\begin{equation}
\int_{0}^{\frac{m^{2}}{M_{Z}^{2}}} \hspace{0.5cm} h_{t}^{4} ( Q^{2})
d(\ln\frac{Q^{2}}{M_{Z})^{2}}) <                  
h_{t}^{4}(M_{Z})^{2}) \ln \frac{m^{2}}{M_{Z}^{2}}
\end{equation}
Higgs
boson masses can be larger than it is obtained, when the
renormalization group equations analysis approach is used
(compare the transformations (20) and the renormalization
group equations for $a_{i}$, i=1,...4).
It is clear that such a difference is most significant for
$M_{s} \sim 10 TeV$ (see formulae (25)-(27) in the next chapter).
Such a difference can be obtained also
in  the case of the absence of spontaneous CP-violation,
if the supersymmetry breaking scale $M_{s} \sim 10 TeV$ is considered.
However, for $M_{s} \sim 1 TeV$ this difference is invisible \cite{3,5,6}.
The same result for $M_{s} \sim 1 TeV$ is obtained here in the
case of the presence of spontaneous CP-violation for small values of
$\tan\beta$
(see formula (25) in the
next chapter).
For large values of $\tan \beta$ an additional increasing of masses
of neutral Higgs bosons A, $h_{1}$ and charged Higgs boson $H^{+}$
takes place because now due to the largeness of $h_{b}$ b-quark
loops contribution to above mentioned particles masses is also
significant (it is seen from (19) and (20) the mass of the
Higgs boson $h_{2}$ doesn't depend on the b-quark and squark one-loop
corrections). The same inequality as (21) can be written also for
b-quark coupling constant $h_{b}$. However, it is necessary to
stress that such an inequality for $h_{b}$ is much stronger
because of $m_{b} \ll M_{Z}$ so that for large values of
$\tan \beta$ the above mentioned increasing of Higgs boson
masses is visible also for $M_{s} \sim 1 TeV$ (see formulae
(28)-(29) in the next chapter).
Summarizing the above discussion, we come to conclusion that
in the case of the presence of spontaneous CP-violation the
restrictions on Higgs boson masses is less severe, if an effective
potential approach applies.

8.Let us now consider the numerical results, which are obtained, if
an effective potential approach, as more general method, applies.
As we had mentioned above, in this case the
lightest Higgs boson must be almost $SU(2)_{L} \times U(1)_{Y}$ singlet
to avoid the contradiction with experiment (as the numerical investigations
show the restriction on the lightest neutral Higgs $m_{h} < 35 GeV$
remain valid, if an effective potential method applies).
During the numerical
investigations two cases were considered: first, when the lightest
neutral Higgs boson is at least $90\%$ singlet (case I)
and the second, when
it is at least $99\%$ singlet (case II). For the case I we
obtain that the lightest Higgs boson has mass
$m_{h_{s}} < 0.5 GeV, 1GeV, 9GeV$ for $M_{s} \sim 100GeV, 1TeV, 10TeV$
respectively.
For the case II we find that
$m_{h_{s}} < 0.2 GeV, 0.8GeV, 3GeV$ for $M_{s} \sim 100GeV, 1TeV, 10TeV$
respectively.
The existence of such a
neutral Higgs bosons with such masses is not excluded by
experiment \cite{16}.

The upper bounds as
functions of $\tan\beta$, obtained for masses of two lightest
nonsinglet mixings of CP-even
fields $\Phi_{1}$, $\Phi_{2}$, $N_{1}$ ($h_{1}$ and $h_{2}$
respectively), for the mass of neutral CP-odd Higgs field
A and for the mass of charged Higgs $H^{+}$ are presented in
Fig. 2,3: in the Fig. 2 for the case I
and in the Fig. 3 for the case II.
As it was expected, the scenario with spontaneous CP-violation
is possible now also for the supersymmetry breaking scale
$M_{s} \sim 100GeV$. However, while for $M_{s} \sim 10TeV$ all
of the values of $1.5 < \tan\beta < 60$ are allowable, for
$M_{s} \sim 1 TeV$ and $M_{s} \sim 100GeV$
the region of values of $\tan\beta$ exists, where
the scenario with spontaneous CP-violation can't be realized.
Besides this, although both in the case I and in the case
II general restrictions on the detectable
(nonsinglet) Higgs particles masses are almost the same (with accuracy of
$10\%$), the restrictions for separate values of $\tan\beta$
in the case II are
much stronger. In particular,
for $M_{s} \sim 1 TeV$ the region of excluded values
of $\tan\beta$ in the case II is larger. The main cause
of such  a difference is the existence in the case I
of additional allowable area of parameter
space of theory: $\varphi,\alpha \ll 1$,
$M_{35}^{2} < \sqrt{10} M_{33}^{2}$, which was found numerically.
As it follows from Fig.2,3
\begin{eqnarray}
\nonumber
m_{h_{1}} < 80 GeV, \hspace{0.5cm}
m_{A} < 110 GeV,                  \\
m_{h_{2}} < 105 GeV , \hspace{0.5cm}                      
m_{H^{+}} < 110 GeV
\end{eqnarray}
for $M_{s} \sim 100 GeV$,
\begin{eqnarray}
\nonumber
m_{h_{1}} < 90 GeV, \hspace{0.5cm}
m_{A} < 115 GeV,     \\                                 
m_{h_{2}} < 140 GeV , \hspace{0.5cm}
m_{H^{+}} < 140 GeV
\end{eqnarray}
for $M_{s} \sim 1 TeV$,
\begin{eqnarray}
\nonumber
m_{h_{1}} < 135 GeV, \hspace{0.5cm}
m_{A} < 140 GeV,       \\                               
m_{h_{2}} < 180 GeV, \hspace{0.5cm}
m_{H^{+}} < 185 GeV
\end{eqnarray}
for $M_{s} \sim 10 TeV$.
It is interesting to compare the numerical results, obtained
using the renormalization group equations analysis approach
(method 1)
with the results, obtained above (method 2). We do this for two cases:
for low values of $\tan\beta$ ($\tan\beta < 10$)
 and for large values
of $\tan\beta$ ($\tan\beta > 20$).
For low values of $\tan\beta$
\begin{eqnarray}
\nonumber
m_{h_{1}} < 90 GeV, \hspace{0.5cm}
m_{A} < 115 GeV,     \\                                 
m_{h_{2}} < 140 GeV , \hspace{0.5cm}
m_{H^{+}} < 90 GeV
\end{eqnarray}
for $M_{s} \sim 1 TeV$, both of approaches
\begin{eqnarray}
\nonumber
m_{h_{1}} < 95 GeV, \hspace{0.5cm}
m_{A} < 125 GeV,       \\                               
m_{h_{2}} < 155 GeV, \hspace{0.5cm}
m_{H^{+}} < 90 GeV
\end{eqnarray}
for $M_{s} \sim 10 TeV$, method 1,
\begin{eqnarray}
\nonumber
m_{h_{1}} < 105 GeV, \hspace{0.5cm}
m_{A} < 130 GeV,       \\                               
m_{h_{2}} < 175 GeV, \hspace{0.5cm}
m_{H^{+}} < 95 GeV
\end{eqnarray}
for $M_{s} \sim 10 TeV$, method 2.

For large values of $\tan\beta$
\begin{eqnarray}
\nonumber
m_{h_{1}} < 60 GeV, \hspace{0.5cm}
m_{A} < 65 GeV,     \\                                 
m_{h_{2}} < 135 GeV , \hspace{0.5cm}
m_{H^{+}} < 115 GeV
\end{eqnarray}
for $M_{s} \sim 1 TeV$, method 1
\begin{eqnarray}
\nonumber
m_{h_{1}} < 85 GeV, \hspace{0.5cm}
m_{A} < 90 GeV,     \\                                 
m_{h_{2}} < 135 GeV , \hspace{0.5cm}
m_{H^{+}} < 140 GeV
\end{eqnarray}
for $M_{s} \sim 1 TeV$, method 2
\begin{eqnarray}
\nonumber
m_{h_{1}} < 90 GeV, \hspace{0.5cm}
m_{A} < 95 GeV,       \\                               
m_{h_{2}} < 155 GeV, \hspace{0.5cm}
m_{H^{+}} < 145 GeV
\end{eqnarray}
for $M_{s} \sim 10 TeV$, method 1,
\begin{eqnarray}
\nonumber
m_{h_{1}} < 135 GeV, \hspace{0.5cm}
m_{A} < 140 GeV,       \\                               
m_{h_{2}} < 180 GeV, \hspace{0.5cm}
m_{H^{+}} < 185 GeV
\end{eqnarray}
for $M_{s} \sim 10 TeV$, method 2.

It is seen that the results, represented by (25)-(31), are in
agreement with the predictions, done in previous chapter.

Summarizing the above discussion, we must also note that the restrictions
on Higgs boson masses, obtained here, are much stronger, compared with
ones, obtained in \cite{3}.

9.Thus we have shown that the scenario with spontaneous CP- violation in
NMSSM to
be real, the lightest neutral Higgs have to be almost $SU(2) \times U(1)$
singlet. This particle  has mass of the order of a few GeV or smaller.
This possibility is not excluded by experiment.
The charged Higgs boson has mass of the order of $\eta$ or smaller
and at least three detectable (i.e. nonsinglet) neutral Higgs
bosons exist with masses of the same order.

So our analysis of the problem of Higgs masses in NMSSM
with the spontaneous CP-breaking shows that the considering model
leads to the predictions for Higgs particles masses, which can be
verified experimentally in the near future.

   The research described in this publication was made possible in part
by Grant No MVU000 from the International Science Foundation.

One of authors (Asatrian H. M.) wants to thank ICTP High Energy Section,
where this work was completed, for hospitality.

\newpage
\begin{figure}[htb]
\epsfxsize=15cm
\epsfysize=18cm
\mbox{\hskip -1.0in}\epsfbox{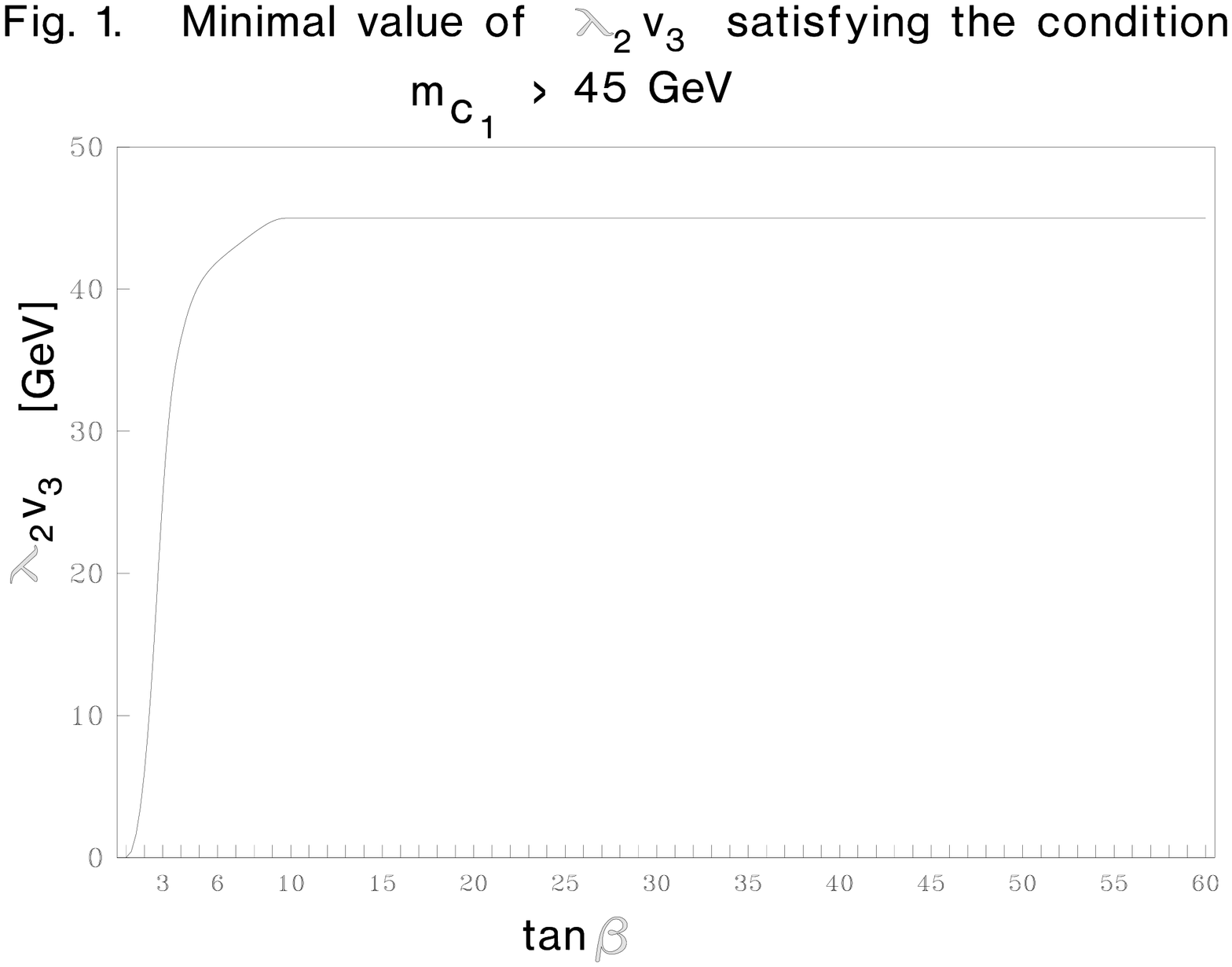}
\end{figure}
\begin{figure}[htb]
\epsfxsize=15cm
\epsfysize=6cm
\mbox{\hskip -1.0in}\epsfbox{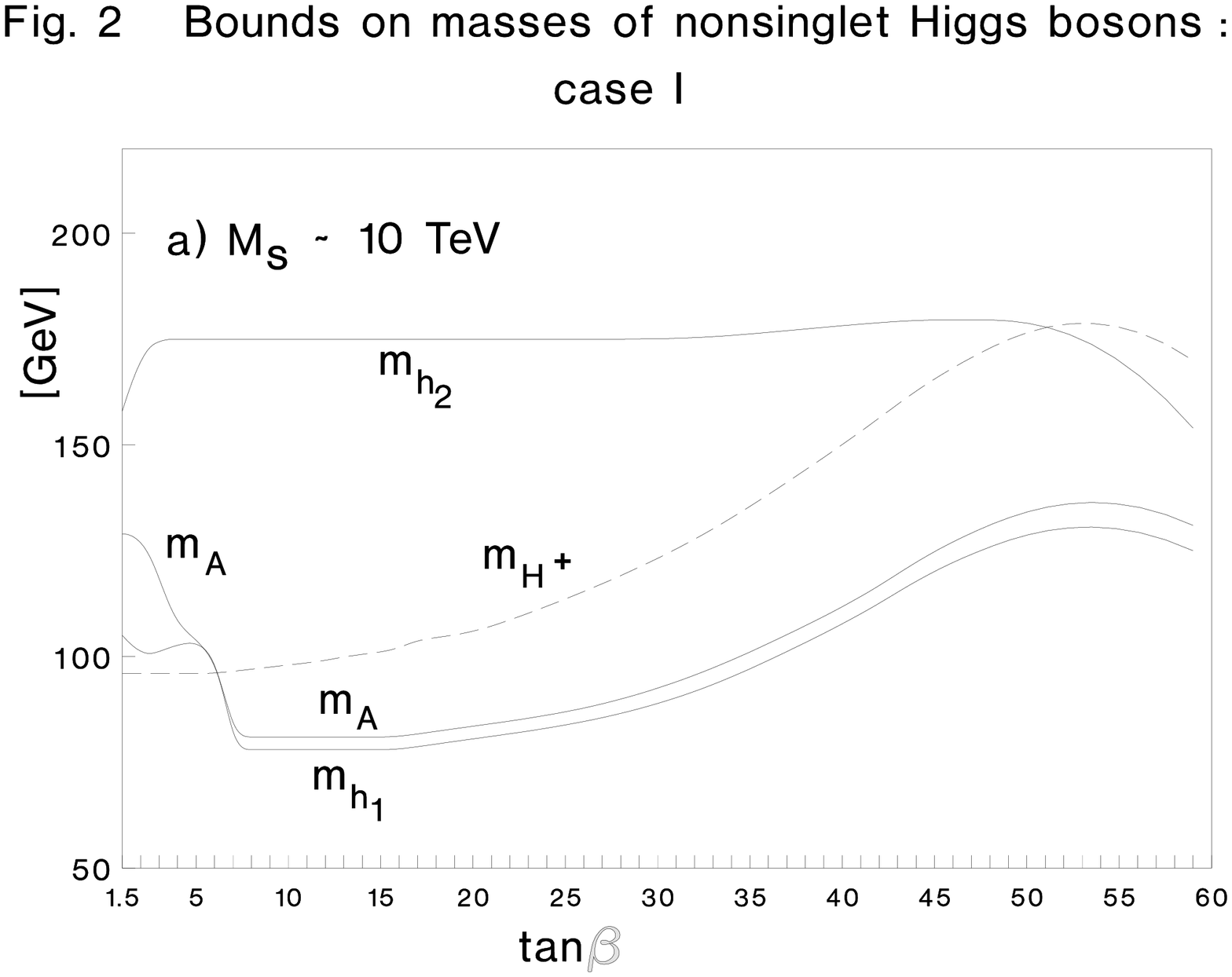}
\end{figure}
\begin{figure}[htb]
\epsfxsize=15cm
\epsfysize=6cm
\mbox{\hskip -1.0in}\epsfbox{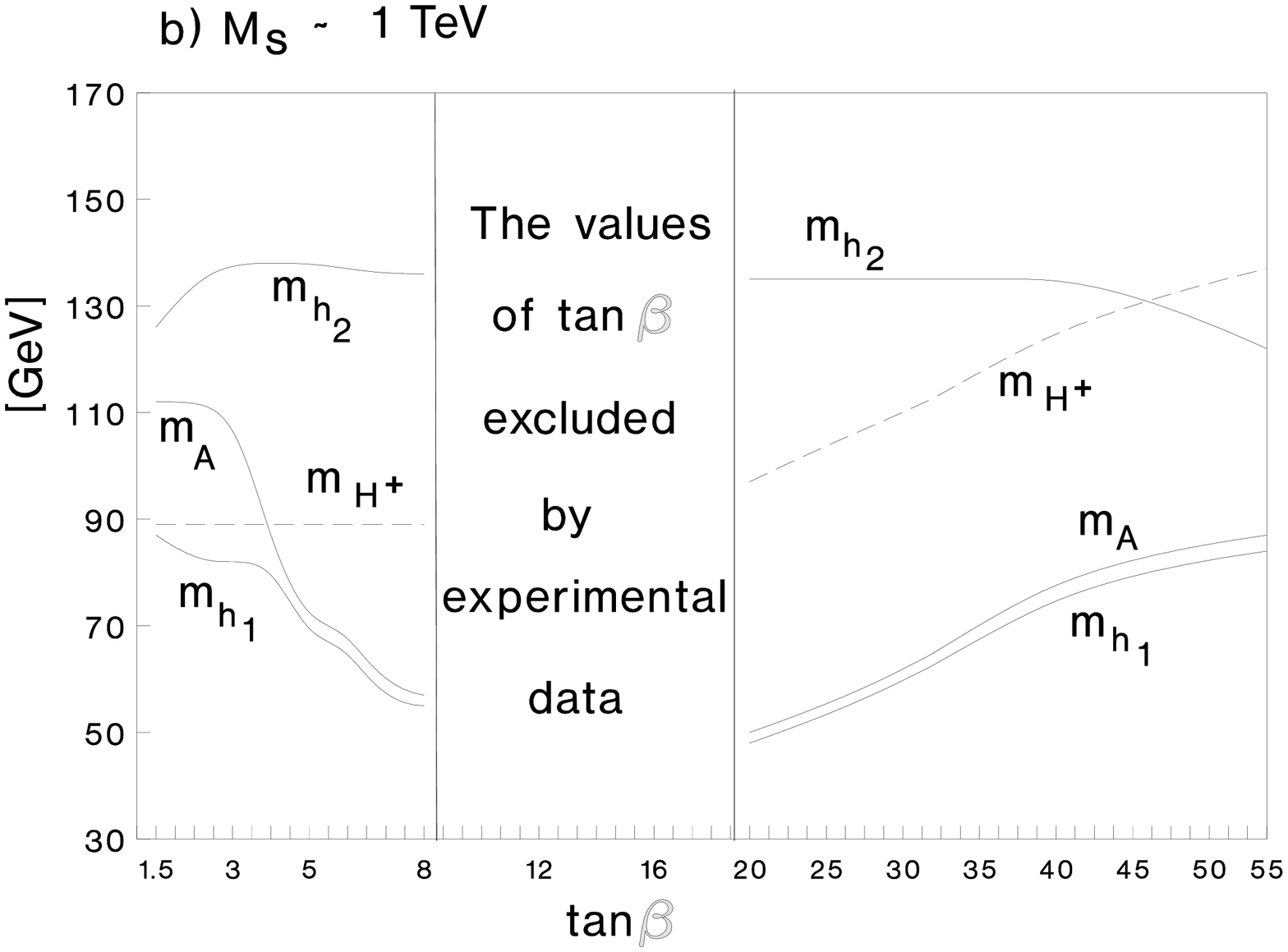}
\end{figure}
\begin{figure}[htb]
\epsfxsize=15cm
\epsfysize=6cm
\mbox{\hskip -1.0in}\epsfbox{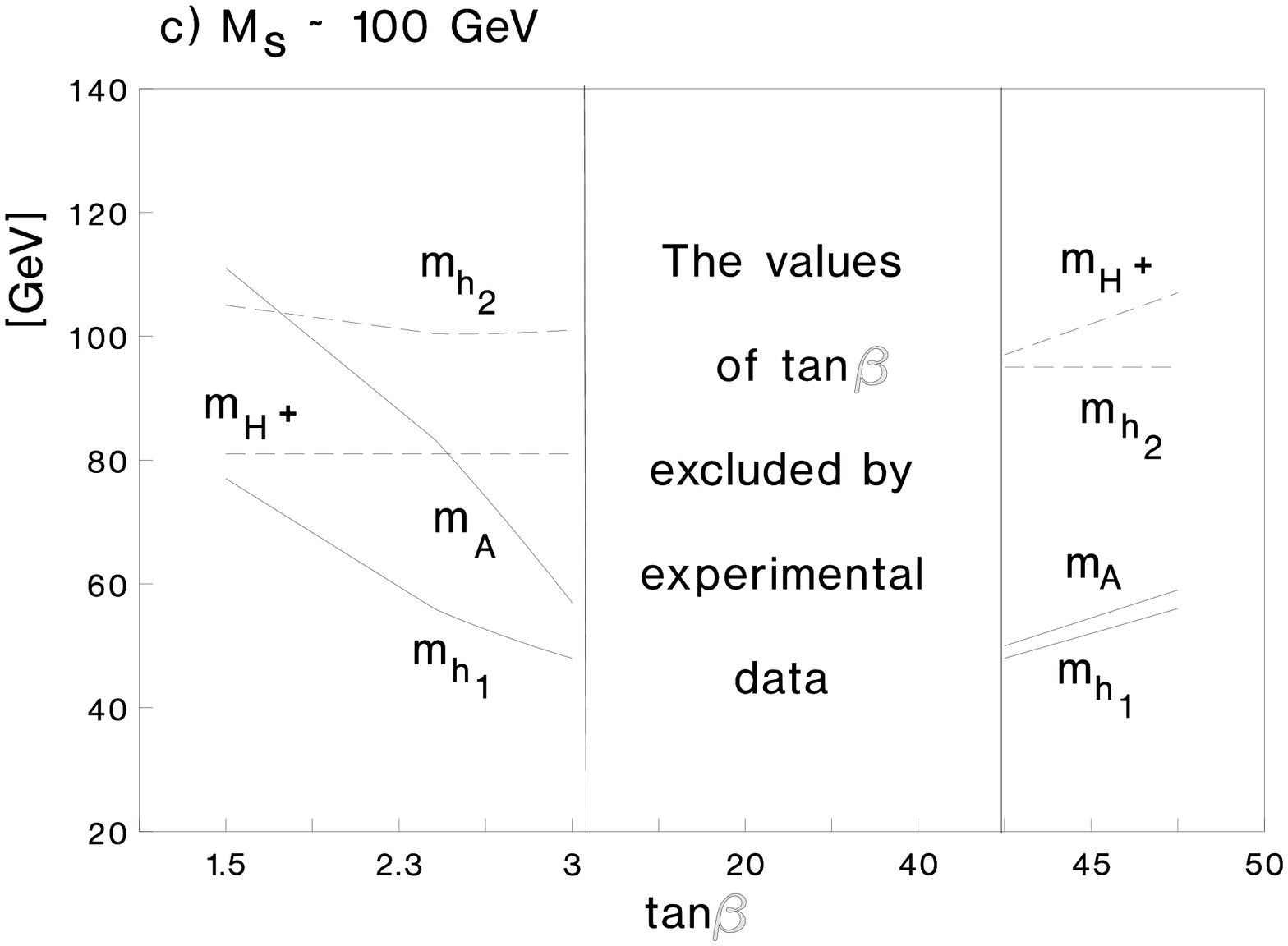}
\end{figure}
\begin{figure}[htb]
\epsfxsize=15cm
\epsfysize=6cm
\mbox{\hskip -1.0in}\epsfbox{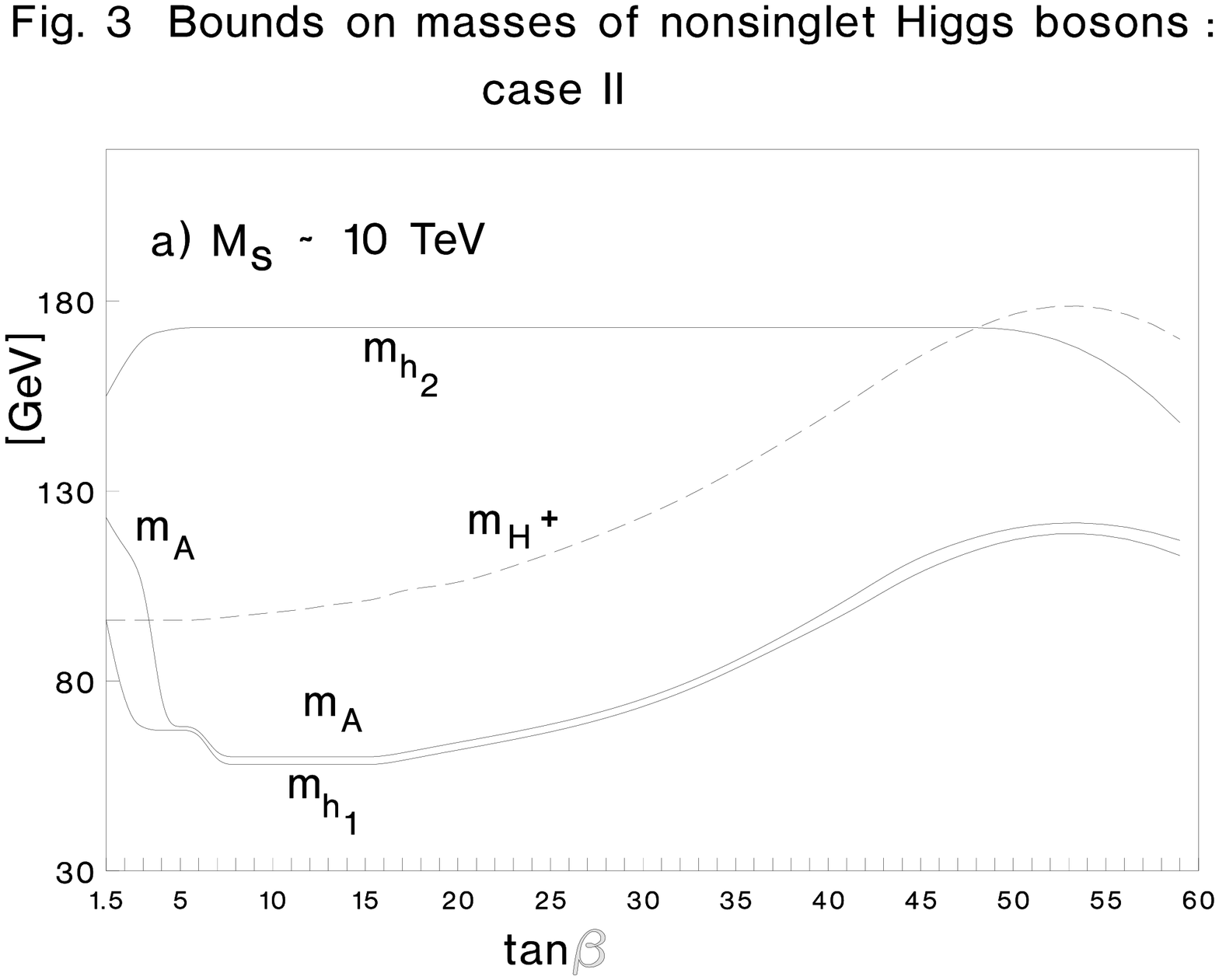}
\end{figure}
\begin{figure}[htb]
\epsfxsize=15cm
\epsfysize=6cm
\mbox{\hskip -1.0in}\epsfbox{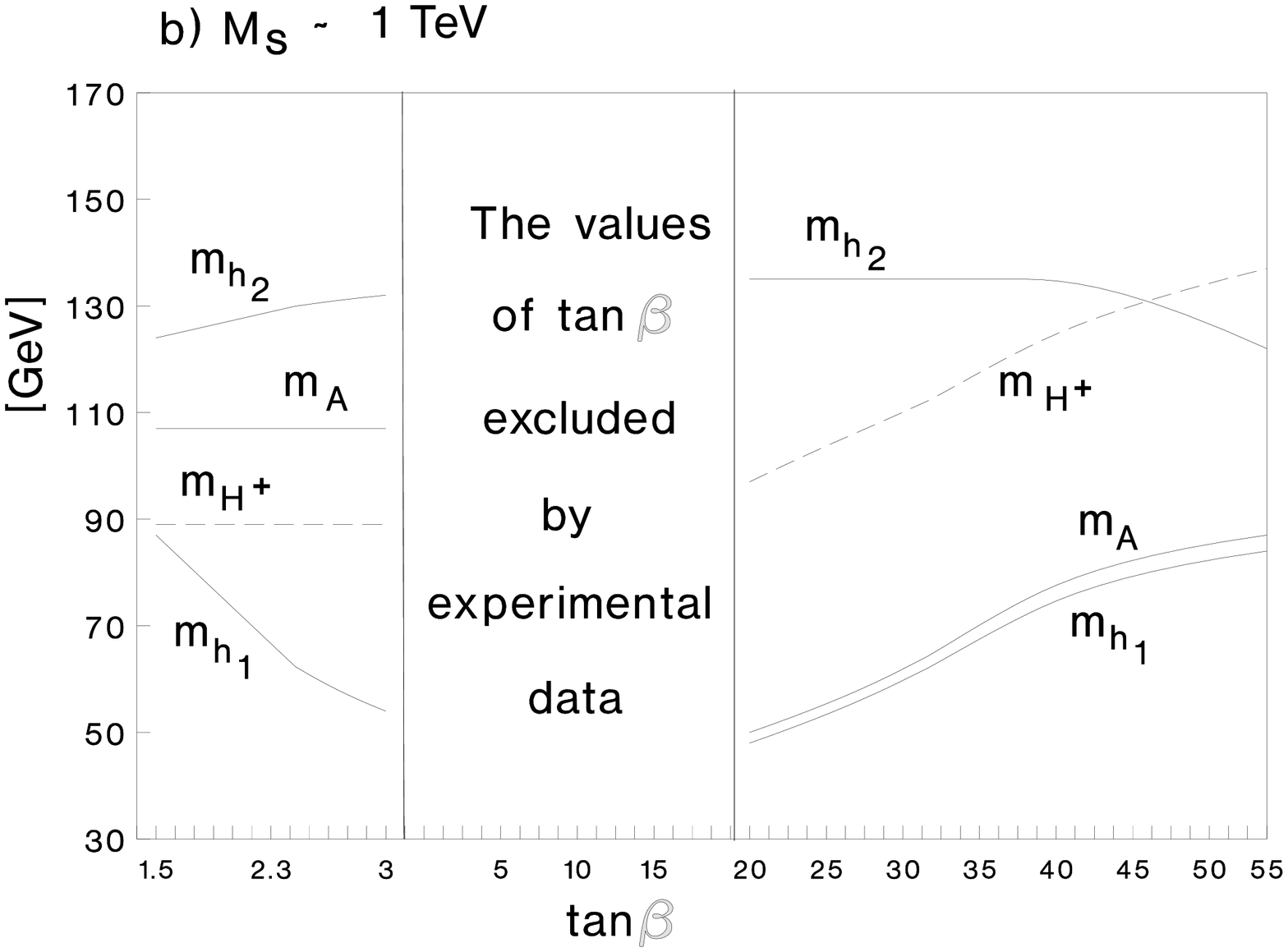}
\end{figure}
\begin{figure}[htb]
\epsfxsize=15cm
\epsfysize=6cm
\mbox{\hskip -1.0in}\epsfbox{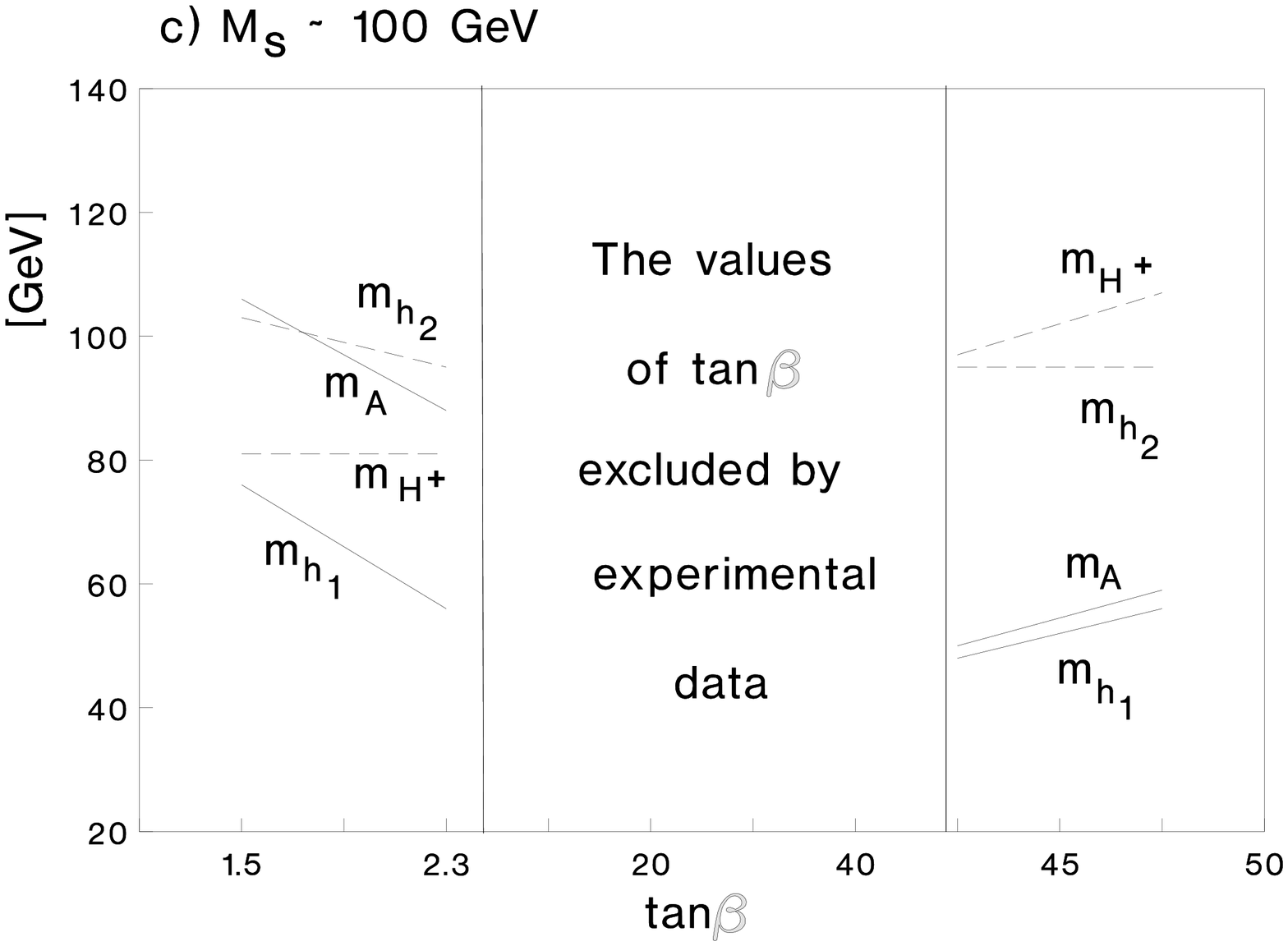}
\end{figure}
\end{document}